\documentclass[11pt]{article}

\usepackage{latexsym}
\usepackage[dvips]{graphicx}

\usepackage{amsmath,amssymb}

\newtheorem{thm}{Theorem}[section]
\newtheorem{df}[thm]{Definitions}
\newtheorem{lm}[thm]{Lemma}

\newcommand{\qed}{\hspace*{\fill} $\Box$}

\title{Every property is testable on a natural class of scale-free multigraphs --- ver.~2}

\author{
Hiro Ito\thanks{
     School of Informatics and Engineering, 
    The University of Electro-Communications (UEC), 
    Tokyo, Japan; and
    CREST, JST, Tokyo, Japan; 
    {\tt itohiro@uec.ac.jp}
    }
 }

\begin{document}

\setcounter{page}{0}

\maketitle

\begin{abstract}
In this paper, we introduce a natural class of multigraphs called hierarchical-scale-free (HSF) multigraphs, and consider constant-time testability on the class. 
We show that a very wide subclass, specifically, that in which the power-law exponent is greater than two, 
of HSF is hyperfinite. 
Based on this result, an algorithm for a deterministic partitioning oracle can 
be constructed.  
We conclude by showing that every property is constant-time testable 
on the above subclass of HSF. 
This algorithm utilizes findings by Newman and Sohler of STOC'11.  
However, their algorithm is based on the bounded-degree model, 
while it is known that actual scale-free networks 
usually include hubs, which have a very large degree.  
HSF is based on scale-free properties and includes such hubs.
This is the first universal result of constant-time testability on  
the general graph model, and it has the potential to be applicable on  a very wide range of scale-free networks. 
\end{abstract}

\newpage
\section{Introduction}


How to handle big data is a very important issue in computer science. 
In the theoretical area, developing efficient algorithms for handling big data 
is an urgent task. 
For this purpose, sublinear-time algorithms look like they could be powerful tools, 
as they are able to read very small parts (constant size) of inputs. 

Property testing is the most well-studied area in sublinear-time algorithms. 
A testing algorithm (or a tester) for a property accepts an input if it has the stipulated property and rejects it if it is far away from having the stipulated property with a high probability (e.g., at least $2/3$) by reading 
a constant part of the input. 
A property is said to be testable if there is a tester~\cite{PropertyTestingLNCS10}.

Property testing of graph properties has been well studied and many fruitful results have been obtained \cite{AlonFNS_Testable_Regularity_SIAMJC09,BSS_MC-testable_STOC08,PropertyTestingLNCS10,GR-STOC97,GGR-JACM98,HKNO_LocalPartition_FOCS09,NS_Testable_SJCOMP13,Levi-Ron_PO_ICALP13}.
Testers on the graphs are separated into three groups according to model: 
the dense-graph model (the adjacent-matrix model), 
the bounded-degree model, 
and 
the general model. 
The dense-graph model is the best clarified: 
In this model, the characteristics of testable properties have been obtained~\cite{AlonFNS_Testable_Regularity_SIAMJC09}.
However, graphs based on actual networks are usually sparse and unfortunately the dense-graph model has been found not to work. 
Studies on the bounded-degree model have been proceeding recently. 
One of the most important findings for this model is that every minor-closed property is testable~\cite{BSS_MC-testable_STOC08}.  
This result can be extended to the surprising result that every property of a hyperfinite graph is testable~\cite{NS_Testable_SJCOMP13}.
However, graphs based on actual models have no degree bounds, i.e., it is known that web-graphs have hubs~\cite{AlbertBarabasi_SF_02,KleinbergLawrence-web-science01}, which have a large degree, and, unfortunately once again, these algorithms do not work for them. 

Typical big-data graph models are scale-free networks, 
which are characterized by the power-law degree distribution.   
Many models have been proposed for scale-free networks \cite{AlbertBarabasi_SF_02,BM_Kronecker_IPL98,Broder_etal-GSinWeb-CN00,CU_SF_Clique_10,Gao_SF_clique_TCS09,KleinbergLawrence-web-science01,StochKronecherG_WAW07,Newman_SF_03,Uno-Watanabe_ScaleFree,WattsStrogatz-Nature98,ZRC_SF_clique_06}. 
Recently, a promising model based on another property of a hierarchical isomorphic structure 
has been presented: 
If we look at a graph in a broad perspective, 
we find a similar structure to local structures. 
Shigezumi, Uno, and Watanabe~\cite{Uno-Watanabe_ScaleFree}
presented a model that is based on the idea of the hierarchical isomorphic 
structure of power-law distribution of isolated cliques.  
An idea of isolated cliques was given by 
Ito and Iwama~\cite{IsoClique_TALG09,IsoClique_ESA05}, 
and the definition is as follows. 
For a nonnegative integer $c \geq 0$, a {\em $c$-isolated clique}
is a clique such that the number of outgoing edges (edges between the clique and the other vertices) 
is less than $ck$, where $k$ is the number of vertices of the clique. 
A 1-isolated clique is sometimes simply called an {\em isolated clique}. 


Based on the model of \cite{Uno-Watanabe_ScaleFree}, we introduce a class of multigraphs, 
hierarchical scale-free multigraphs (HSF, Definitions~\ref{df:HSF})\footnote{
In a preliminary version of this paper, \cite{ScalefreeTest_itohiro15_arXiv}, 
the definition of HSF is different.  
The definition in this paper is far more general (wider) than in the preliminary version. 
}, 
which represents natural scale-free networks. 
We show the following result (Theorem~\ref{th:testable}): 

\begin{quote}
{\em Every property is testable on HSF if the power-law exponents are greater than two}.  
\end{quote}

Given this result, many problems on actual scale-free big networks will prove to be solvable in constant time. 
Although this result is an application of the algorithms of \cite{NS_Testable_SJCOMP13},  
which is a result on bounded-degree graphs, 
HSF is not a class of bounded-degree graphs. 
This is the first result on universal algorithms for the general graph model.

\subsection{Definitions}

In this paper, we consider undirected multigraphs without self-loops.  
We simply call this type of multigraph a ``graph'' in this paper 
and use $G = (V,E)$ to denote it, 
where $V$ is the vertex set and $E$ is the edge (multi)set. 
Sometimes $V$ and $E$ are denoted by $V[G]$ and $E[G]$, respectively. 
Henceforth, we use ``set'' to refer to a multiset for notational simplicity. 
Throughout this paper, $n$ is used to denote the number of vertices of 
a graph, i.e., $|V| = n$.

For a graph $G = (V,E)$ and vertex subsets $X,Y \subseteq V$, 
 $E_G(X,Y)$ denotes the edge set between $X$ and $Y$, i.e., 
 $E_G(X,Y) = \{ (x,y) \in E ~|~ x \in X, y \in Y \}$. 
 $E_G(X,V \backslash X)$ is also simply written as $E_G(X)$. 
$|E_G(X)|$ is denoted by $d_G(X)$. 
For a vertex $v \in V$, the number of edges incident to $v$ is called the {\em degree} of $v$.  
A singleton set $\{ x \}$ is often written as $x$ for notational simplicity. 
E.g., the degree of $v$ is represented by $d_G(v)$.   
The subscript $G$ in the above $E_G(*)$, $d_G(*)$, etc., 
may be omitted if it is clear. 

For a vertex $v \in V$, $\Gamma_G(v)$ denotes the set of vertices adjacent to $v$, 
i.e., $\Gamma_G(v) := \{ u \in V ~|~ (v,u) \in E \}$. 
Note that $|\Gamma_G (v)|$ may not be equal to $d_G(v)$ 
as parallel edges may exist. 
For a graph $G = (V,E)$ 
and a vertex subset $X \subseteq V$, 
the {\em subgraph induced by $X$} is 
defined as $G(X) = (X, \{ (u,v) \in E ~|~ u,v \in X \} )$.

For a vertex subset $X \subseteq V$, 
a {\em contraction} of $X$ is defined as 
an operation to 
(i) replace $X$ with a new vertex $v_X$, 
(ii) replace each edge $(v,u)$ in $E(X)$ ($v \in X, u \in V \backslash X$) with 
a new edge $(v_X,u)$, and 
(iii) remove all edges between vertices in $X$. 
That is, by contracting $X \subseteq V$, 
a graph $G=(V,E)$ is changed to $G'=(V',E')$ such that 
\begin{eqnarray*}
&V' = V \backslash X \cup \{ v_X \}, ~\mbox{and}& \\
&E' = E \backslash \{ (v,u) ~|~ v \in X, u \in V \} \cup 
\{ (v_X,u) ~|~ (v,u) \in E, v \in X, u \in V-X \}. &
\end{eqnarray*}
We identify the above $(v_X,u) \in E'$ with $(v,u) \in E$. 
In other words, we say that $(v,u)$ remains in $G'$ (as $(v_X,u)$). 
Note that the graphs are multigraphs, and thus if there are two edges $(v,u),(v',u) \in E$ for $v,v' \in X$, $v \neq v'$ and $u \in V \backslash X$, 
then two parallel edges, both represented by $(v_X,u)$, 
one of which corresponds to $(v,u)$ and 
the other of which corresponds to $(v',u)$, are added to $E'$. 
Also note that none of the graphs considered in this paper contain self-loops, and hence 
an edge $(v,v') \in E$ with $v,v' \in X$ is removed by contracting $X$.

Two graphs $G_1 =(V_1, E_1)$ and $G_2 =(V_2, E_2)$ 
are {\em isomorphic} if there is a one-to-one correspondence 
$\Phi: V_1 \to V_2$ 
such that $E_{G_1}(u,v) = E_{G_2}(\Phi(u), \Phi(v))$ 
for all $u,v \in V_1$. 
A graph property (or property, for short) is a (possibly infinite) 
family of graphs,  
which is closed under isomorphism.

\begin{df}[$\epsilon$-far and $\epsilon$-close]
Let $G =(V, E)$ and $G' =(V', E')$ be two graphs 
with $|V| = |V'| =n$ vertices. 
Let $m(G,G')$ be the number of edges 
that need to be deleted and/or 
inserted from $G$ in order to make it isomorphic to $G'$. 
The distance between $G$ and $G'$ is defined as\footnote{
The distance defined here may be larger than 1 
as $m(G,G') > n$ may occur. 
(In the bounded-degree model 
it is defined as $\mbox{\rm dist}(G,G') = m(G,G')/dn$.) 
However, here we consider sparse graphs 
and they have an implicit upper bound of the average (not possibly maximum) degree, 
say $d$, and thus $\mbox{\rm dist}(G,G')$ is bounded by $d$. 
} 
$\mbox{\rm dist}(G,G') = m(G,G')/n$. 
We say that 
$G$ and $G'$ are {\em $\epsilon$-far} if 
$\mbox{\rm dist}(G,G) > \epsilon$; 
otherwise {\em $\epsilon$-close}. 
Let $P$ be a non-empty property. 
The distance between $G$ and $P$ is 
$\mbox{\rm dist}(G,P) = \min_{G'' \in P} \mbox{\rm dist}(G,G'')$. Otherwise we say that 
$G$ is {\em $\epsilon$-far} from $P$ if $\mbox{\rm dist}(G,P) > \epsilon$, 
and {\em $\epsilon$-close}. 
\end{df}

\begin{df}[testers]
A {\em testing algorithm} for a property $P$ is an algorithm that, given query access 
to a graph $G$, accepts every graph from $P$ 
with a probability of at least $2/3$, and rejects every graph that is $\epsilon$-far 
from $P$ with probability at least $2/3$. 
Oracles in the general graph model are: for any vertex $v$, the algorithm may ask for the degree $d(v)$, and may ask for the $i$th neighbor of the vertex (for $1 \leq i \leq d(v)$).\footnote{Although asking whether there is an edge between any two vertices is also allowed in the general graph model, the algorithms we use in this paper do not need to use this query.} 
The number of queries made by an algorithm to the given oracle is called 
the {\em query complexity} of the algorithm. 
If the query complexity of a testing algorithm is a constant, 
independent of $n$ (but it may depend on $\epsilon$), 
then the algorithm is called a {\em tester}. 
A (graph) property is {\em testable} if there is a tester for the property. 
\end{df}

\begin{df}[isolated cliques~\cite{IsoClique_TALG09}]
For a graph $G = (V,E)$ and a real number $c \geq 0$, 
a vertex subset $Q \subseteq V$ is called a {\em $c$-isolated clique} 
if $Q$ is a clique (i.e., 
$(u,v) \in E$, for all $u,v \in Q$ and $u \neq v$) 
and $d_G(Q) < c |Q|$. 
A 1-isolated clique is sometimes called an {\em isolated clique}.
${\cal E}(G)$ is the graph obtained from $G$ by contracting all isolated cliques. 
Two distinct isolated cliques never overlap, 
except in the special case of {\em double-isolated-cliques}, which consists 
of two isolated cliques with size $k$ sharing $k-1$ vertices.  
A double-isolated-clique $Q$ has no edge between $Q$ and the 
other part of the graph (i.e., $d_G(Q)=0$), and thus we specially define that a double-isolated-clique in $G$ is contracted into  
a vertex in ${\cal E}(G)$. 
Under this assumption, ${\cal E}(G)$ is uniquely defined. 
\end{df}

\begin{df}[hyperfinite~\cite{hyperfinite}]
For real numbers $t >0$ and $\epsilon >0$, 
a graph $G=(V,E)$ consisting of $n$ vertices is {\em $(t,\epsilon)$-hyperfinite} 
if one can remove at most $\epsilon n$ edges from $G$ 
and obtain a graph whose connected components have size at most $t$. 
For the function $\rho: \mbox{\boldmath$R$}^+ \to \mbox{\boldmath$R$}^+$, 
$G$ is {\em $\rho$-hyperfinite} 
if it is $(\rho(\epsilon),\epsilon)$-hyperfinite for all $\epsilon >0$. 
A family ${\cal G}$ of graphs is {\em $\rho$-hyperfinite}  
if all $G \in {\cal G}$ are $\rho$-hyperfinite. 
A family ${\cal G}$ of graphs is {\em hyperfinite} 
if there exists a function $\rho$ such that ${\cal G}$ is $\rho$-hyperfinite. 
\end{df}

Hyperfinite is a large class, as it is known that any minor-closed property is hyperfinite
in a bounded-degree model. From the viewpoint of testing, the importance of hyperfiniteness stems from the following result.

\begin{thm}[\cite{NS_Testable_SJCOMP13}]
For the bounded-degree model, any property is testable for any class of hyperfinite graphs. 
\end{thm}


This result is very strong, but there is a problem in that the result works on bounded-degree graphs and it is natural to consider that actual scale-free networks do not have a degree bound.

\subsection{Our contribution and related work}

In this paper, we apply the universal algorithm of \cite{NS_Testable_SJCOMP13} to scale-free networks. 
We formalize two natural classes, 
${\cal SF}$ and ${\cal HSF}$
that represent scale-free networks\footnote{
${\cal HSF}$ was introduced in the preliminary version of this paper \cite{ScalefreeTest_itohiro15_arXiv}. 
However, the definition in this paper is more general (wider) than in the preliminary version. 
}. 
The latter is a subclass of the former.

\begin{df}\label{df:DSF}
For positive real numbers $c >1$ and $\gamma >1$, 
a class of {\em scale-free graphs (SF)} 
${\cal SF}(c,\gamma)$
consists of (multi)graphs $G=(V,E)$ for which the following condition holds: 
\begin{itemize}
\item[(i)] Let $\nu_i$ be the number of vertices $v$ with $d(v) = i$. Then: 
\begin{equation}
\nu_i \leq  c n i^{- \gamma}, 
~~~ \forall i \in \{ 2, 3, \ldots, \}. \label{eq:DSF}
\end{equation}
\end{itemize}
\end{df}

The above property (i) is generally called a power-law and in many actual scale-free networks, it is said that $2 < \gamma < 3$~\cite{AlbertBarabasi_SF_02}. 
That is, ${\cal SF}$ is a class of multigraphs that obey the power-law degree distribution.

We show that this class is $\epsilon$-close to a bounded-degree class if $\gamma >2$ (Lemma~\ref{lm:degreebound}). 

After showing this property, we show the hyperfiniteness of the class. 
Hyperfiniteness seems to be closely related to a high clustering coefficient, where 
the cluster coefficient $\mbox{cl}(G)$ of a graph $G =(V,E)$ 
is defined as\footnote{
There is another way to define the cluster coefficient: 
$3 \times (\mbox{\# of cycles of length three}) / (\mbox{\# of paths of length two})$. 
Although these two values are different generally, they are close under the assumption of the power-law degree distribution.  
}:
$$
\mbox{cl}(G) := \frac{1}{n} \sum_{v \in V} \mbox{cl}_G(v),  
~~~~ \mbox{cl}_G(v) := \frac{|\{ (u,v) \in E ~|~ u,v \in \Gamma_G (v), u \neq v \}|}{{|\Gamma_G (v)| \choose 2}}
$$

Sometimes $\mbox{cl}_G(v)$ is called the {\em local cluster coefficient} of $v$. 
It is said that $\mbox{cl}(G)$ is $O(1)$ for 
many classes that model actual social networks, while 
$\lim_{n \to \infty} \mbox{cl}(G) = 0$ for random graphs. 

These three characterizations, 
``high clustering coefficient,'' 
``existence of isolated cliques,'' 
and 
``hyperfiniteness''  
appear to be closely related to each other.  
In fact, it is readily observed that 
if $\mbox{cl}_G(v) = 1$ for a bounded-degree graph $G$ (the degree bound is $d$), 
then $G$ consists of only (completely) isolated cliques with size at most $d+1$, and 
$G$ is $(d+1, 0)$-hyperfinite! 

Unfortunately, however, 
it is also observed that 
for any $0 < c < 1$, 
there is a class of bounded-degree graphs $G$ 
such that $\lim_{n \to \infty} \mbox{cl}(G) = c$ and 
it is not $(t, \epsilon)$-hyperfinite 
for any pair of constants $t$ and $\epsilon < 1/2$, e.g., 
$G =(V,E)$ 
consists of 
$n/d$ cliques of size $d$, 
and random $n/2$ edges between 
vertices in different cliques (each vertex has $d-1$ adjacent vertices in its clique and one adjacent vertex outside the clique).  
To separate this graph into constant-sized connected components, almost all of the edges between cliques (their number is $\epsilon n/2$) must be removed. 

However, we do not need to give up here, as the above model does not look like a natural model of 
scale-free networks, e.g., by contracting each isolated clique, it becomes a mere random graph with $n/d$ vertices.  
From this fact, the hierarchical structure of a high cluster coefficient looks important. 
The model presented by \cite{Uno-Watanabe_ScaleFree} has such a structure.  
Based on this model, we present the following class of multigraphs:

\begin{df}[Hierarchical Scale-Free Graphs]\label{df:HSF}
For positive real numbers $c, \gamma > 1$ and a positive integer $n_0 \geq 1$, 
a class of {\em hierarchical scale-free graphs (HSF)} ${\cal HSF} = {\cal HSF}(c,\gamma, n_0)$
consists of (multi)graphs $G=(V,E)$ 
for which the following conditions hold: 
\begin{itemize}
\item[(i)] $G \in {\cal SF}(c,\gamma)$
\item[(ii)] Consider the infinite sequence of graphs 
$G_0 =G$, $G_1 = {\cal E}(G_{0})$, $G_2 = {\cal E}(G_{1})$, $\ldots$.   
If $|V[G_i]| \geq n_0$, then $G_i$ includes at least one isolated clique 
$Q \subseteq V$ with $|Q| \geq 2$. 
(Note that if $G_k$ has no such isolated clique, then $G_k = G_{k+1} = G_{k+2} =\cdots$.)
\end{itemize}
\end{df}

We show the following results.

\begin{thm}\label{th:hyperfinite}
For any 
${\cal HSF} = {\cal HSF}(c,\gamma,n_0)$ 
with $\gamma >2$ 
and any real number $\epsilon >0$, 
there is a real number $t_{\mbox{\scriptsize \ref{th:hyperfinite}}}
=t_{\mbox{\scriptsize \ref{th:hyperfinite}}} ({\cal HSF} ,\epsilon)$ 
such that 
${\cal HSF}$ is 
$(t_{\mbox{\scriptsize \ref{th:hyperfinite}}}, \epsilon)$-hyperfinite. 
\end{thm}

We give a global algorithm for obtaining the partition realizing the hyperfiniteness of Theorem~\ref{th:hyperfinite}.  
The algorithm is deterministic, i.e., 
if a graph and the parameter $\epsilon$ are fixed, 
then the partition is also fixed. 
The algorithm can be easily revised to a local algorithm and we obtain a deterministic partitioning oracle to get the partition (Lamma~\ref{lm:PO}).  
Note that all known partitioning algorithms
are randomized algorithms.  
By using this partitioning oracle and an argument similar to one used in \cite{NS_Testable_SJCOMP13}, we get the following main theorem.

\begin{thm}\label{th:testable}
Any property is testable 
for ${\cal HSF}(c,\gamma,n_0)$ 
with $\gamma >2$. 
\end{thm}

\paragraph{Related work}
As stated earlier, for the bounded-degree model, Newman and Sohler~\cite{NS_Testable_SJCOMP13} presented a universal tester (which can test any property) 
for hyperfinite graphs.  
In the general graph model, there exist far fewer results than for the bounded-degree graph model and the dense graph model. 
For universal-type sublinear-time algorithms, 
Kusumoto and Yoshida~\cite{Kusumoto-Yoshida-ICALP14} gave 
a testing algorithm with $\mbox{plylog}(n)$ query complexity for 
forest-isomorphism. In the same paper, they showed an $\Omega (\log n)$ lower bound for this problem.  

This paper gives a universal tester that can test every property on a natural class of scale-free multigraphs in constant time. 
This is the first result for universal constant-time algorithms on sparse and degree-{\em un}bounded graphs.

\section{Hyperfiniteness and a Global Partitioning Algorithm}

\subsection{Degree bounding}

For a graph $G$ and a nonnegative integer $d \geq 0$, 
$G|d$ is a graph made by deleting all edges incident to each vertex $v$ with $d(v) > d$ from $G$. 
Note that $G|d$ is a bounded-degree graph with degree bound $d$. 

\begin{lm}\label{lm:degreebound}
For any ${\cal SF} = {\cal SF}(c,\gamma)$ with $\gamma >2$, 
and any positive real number $\epsilon >0$, 
there is a constant 
$\delta_{\scriptsize \ref{lm:degreebound}} 
= \delta_{\scriptsize \ref{lm:degreebound}}(\epsilon, c,\gamma)$ 
such that for any graph $G \in {\cal SF}$, 
$G|\delta_{\scriptsize \ref{lm:degreebound}}$ is $\epsilon$-close to $G$. 
\end{lm}

Before showing a proof of this lemma, we introduce some definitions. 
Riemann zeta function is defined by
$\zeta (\gamma) = \sum_{i=1}^{\infty} i^{-\gamma}$. 
This function is known to converge to a constant ($\zeta (\gamma) < 1 + (\gamma -1)^{-1}$) 
for any $\gamma >1$. 
We introduce a generalization of this function by using a positive integer $k \geq 1$ as 
$\zeta (k, \gamma) = \sum_{i=k}^{\infty} i^{-\gamma}$. 
Note that $\zeta (\gamma) = \zeta (1, \gamma)$.

\begin{lm}\label{lm:converge_zeta}
For any $\epsilon >0$ and $\gamma >1$, there is an integer 
$k_{\scriptsize \ref{lm:converge_zeta}} 
= k_{\scriptsize \ref{lm:converge_zeta}} (\epsilon, \gamma) \geq 1$ 
such that $\zeta (k_{\scriptsize \ref{lm:converge_zeta}}, \gamma) < \epsilon$. 
\end{lm}

\noindent
{\em Proof}: 
It is clear from the above fact that $\zeta (\gamma)$ converges for every $\gamma >1$. \qed\\

\noindent
{\em Proof of Lemma~\ref{lm:degreebound}}: 
Let $d$ be an arbitrary positive integer. 
Let $m_d$ be the number of removed edges to make $G|d$ from $G$. 
From (\ref{eq:DSF}), 
$$
m_d = \sum_{i=d+1}^{\infty} i \nu_i  
\leq \sum_{i=d+1}^{\infty} c n i^{-(\gamma -1)} 
= c n \zeta (d+1, \gamma -1). 
$$
From the assumption of $\gamma>2$ and Lemma~\ref{lm:converge_zeta}, 
$\zeta (d+1, \gamma -1) < \epsilon/c$ if 
$d +1 \geq k_{\scriptsize \ref{lm:converge_zeta}} (\epsilon/c, \gamma -1)$. 
Thus by letting 
$\delta 
= \delta_{\scriptsize \ref{lm:degreebound}}(\epsilon, c,\gamma) 
= k_{\scriptsize \ref{lm:converge_zeta}} (\epsilon/c, \gamma-1) -1$, 
we have $m_{\delta_{\scriptsize \ref{lm:degreebound}} } < \epsilon n$. \qed\\

From here, we denote the above $\delta_{\scriptsize \ref{lm:degreebound}}(\epsilon, c,\gamma)$ by $\delta$ for notational simplicity.

\subsection{Hierarchical contraction, structure tree, and coloring}

Let $W_1, \ldots, W_k$ ($W_i \subseteq V$, $\forall i \in \{ 1, \ldots, k \}$) 
be a family of subsets of vertices satisfying that 
$W_i \cap W_j = \emptyset$ for every $i,j \in \{ 1, \ldots, k \}$ and $i \neq j$, 
and $W_1 \cup \cdots \cup W_k = V$. 
Then $\{ W_1, \ldots, W_k \}$ is called 
a {\em partition} of $V$. 
Below, we explain a global algorithm for obtaining a partition of $V$ realizing the hyperfiniteness 
of a graph in ${\cal HSF}$ with $\gamma >2$, i.e., 
$|W_i|$ is bounded by a constant and the number of edges between different $W_i$ and $W_j$ is, 
at most, $\epsilon n$. 
First, we give a base algorithm. \\

\noindent
\textbf{procedure} {\sc HierarchicalContraction}($G$)\\
\textbf{begin}\\
1 \hspace{1em} $i :=0$, $G_0:=G$\\
2 \hspace{1em} \textbf{while} 
there exists an isolated clique in $G_i =(V_i,E_i)$ \textbf{do}\\
3 \hspace{2em} $i := i+1$, $G_i:= {\cal E}(G_{i-1})$\\
4 \hspace{1em} \textbf{enddo}\\
\textbf{end.}\\

\begin{figure}[bht]
\begin{center}
\includegraphics[scale=1]{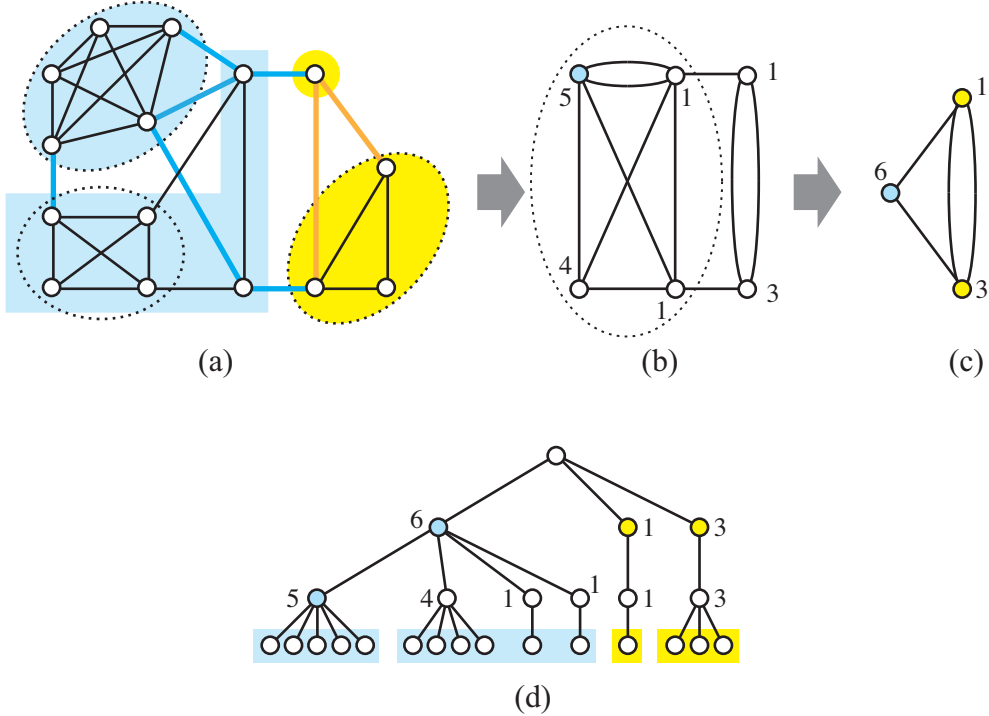}
\end{center}
\caption{An example of {\sc HierarchicalContraction}, the structure tree $T$, and the coloring: 
Here, we assume $\delta/\epsilon = 4.5$; 
the number beside a vertex is $w(*)$; 
the dotted circles are isolated cliques; 
colored areas are blue or yellow components.}
\label{fg:fig-hierarch_contractions}
\end{figure}

We denote $G_i =(V_i,E_i)$ for $i \in \{ 0, 1, \ldots \}$. 
Let $G_k = (V_k,E_k)$ be the final graph of {\sc HierarchicalContraction}($G$). 
From the definitions of HSF, 
$|V_k| < N$. 
See Fig.~\ref{fg:fig-hierarch_contractions}~(a)--(c) for 
an example of applying this procedure. 

The trail of the contraction can be represented by a rooted tree $T = (V[T],E[T])$, 
which is called the {\em structure tree} of $G$,  
defined as follows.  
(Fig.~\ref{fg:fig-hierarch_contractions}~(d) shows 
an example of the structure tree\footnote{
In this example, we ignore $\delta$ and the red vertices. Some may feel it curious that if $\delta / \epsilon = 4.5$, then $\delta << 4.5$ for small $\epsilon$, 
and hence many vertices in this example become red. 
This is correct. However, $w(v)$ becomes larger, step by step, and thus 
if $k$ is very large, then $w(v)$ may be larger 
than $\delta / \epsilon$, by only contracting 
vertices whose degree is at most $\delta$. 
}.)

$V[T] := V_0 \cup V_1 \cup \cdots \cup V_k \cup \{ r \}$, 
where $r$ is the (artificial) root of $T$. 
Each $v \in V_0$ is a leaf of $T$, and 
a vertex $v \in V_i$ ($i \in \{ 0, \ldots, k \}$) is on the level $i$ of $T$, 
i.e., 
$v \in V_i$ ($ i  \geq 1$) is the parent of $u \in V_{i-1}$ 
if ``$v$ is made by contracting a subset (clique or a double-isolated-clique) 
$Q \subseteq V_{i-1}$ such that $u \in Q$'' or  
``$v=u$ (i.e., $u$ is not included in an isolated clique in $G_{i-1}$).''
The root $r$ is the parent of every vertex in $V_k$. 
(The reason $r$ is added is only to make $T$ a tree.)

We introduce a function $W: V[T] \backslash \{ r \} \to 2^{V}$
and coloring on the vertices in $V[T]$ 
as follows: 
\begin{itemize}
  \item For $v \in V_0$: 
  \begin{itemize}
    \item $W(v) = \{ v \}$, and 
    \item if $d(v) > \delta$, then $v$ is colored {\em red}, otherwise uncolored. 
  \end{itemize}
  \item For $v \in V_i$ ($i =1, \ldots, k$): 
  \begin{itemize}
      \item let $S(v)$ be the set of uncolored children of $v$, 
      \item $W(v) = \bigcup_{u \in S(v)} W(u)$, and 
      \item if $|W(v)| > \delta / \epsilon$, then $v$ is colored {\em blue}, 
      \item else if $v \in V_k$ and $W(v) \neq \emptyset$, then $v$ is colored {\em yellow}, 
      \item otherwise, $v$ is uncolored. 
  \end{itemize}
\end{itemize}

Note that for any two distinct colored vertices $u,v \in V[T]$, 
$W(u) \cap W(v) = \emptyset$. 
For every $v \in V[T]$, we also define a weight function as 
$w(v) = |W(v)|$. 
For a blue (resp. yellow) colored vertex $v \in V[T]$, 
$W(v) \subseteq V$ is called a {\em blue (resp. yellow) component}.

By using these colors, 
we also color the edges in $E~(=E_0)$ in the following manner: 
\begin{itemize}
\item For every red vertex $v \in V_0~(=V)$, all edges in $E_G(v)$ are colored {\em red}. 
\item For every blue component $W \subseteq V$, for every edge $e \in E_G(W)$,  
if $e$ is not colored red, then $e$ is colored {\em blue}. 
\item For every yellow component $W \subseteq V$, for every edge $e \in E_G(W)$,  
if $e$ is not colored either red or blue, 
then $e$ is colored {\em yellow}. 
\end{itemize}

The other edges in $E$ are uncolored. 
The set of red, blue, and yellow edges in $E$ 
are represented by $R$, $B$, and $Y$, respectively. 
These colors are preserved in 
$G_1 ={\cal E}(G_0)$, $G_2 ={\cal E}(G_1)$, $\ldots$, $G_k ={\cal E}(G_{k-1})$, 
e.g., if an edge $e \in E_i$ is red, then the corresponding edge in $E_{i+1}$ is also red.

\subsection{Proof of Theorem~\ref{th:hyperfinite}}

Before showing the proof of Theorem~\ref{th:hyperfinite}, we prepare some lemmas.

\begin{lm}\label{lm:high_degree_is_red}
For any $G_i$ ($i \in \{ 0, \ldots, k \}$), 
all edges incident to a vertex with a degree higher than $\delta$ are red.  
\end{lm}

\noindent
{\em Proof}: 
For $G_0 =G$, the statement clearly holds from the coloring rule. 
Assume that the statement holds in $G_{i-1}$, and 
does not hold in some $G_i$. 
Let $v$ be a vertex in $V_{i}$ such that $d_{G_{i}}(v) \geq \delta +1$ 
and a non-red edge is incident to $v$. 
Then $v$ must be made by contracting an 
isolated clique  in $G_{i-1}$, say $Q \subseteq V_{i-1}$,  
such that $d_{G_{i-1}}(Q) \geq \delta+1$. 
From the definition of isolated cliques, 
$|Q| \geq  d_{G_{i-1}}(Q) +1 \geq \delta +2$. 
Since $Q$ is a clique, every vertex $Q$ 
has degree at least $|Q|-1 \geq \delta +1$ in $G_{i-1}$. 
It follows that all edges incident to a vertex in $Q$ must be red. 
This contradicts the assumption that a non-red edge is incident to $v$. \qed

\begin{lm}\label{lm:number_of_colored_edges}
$|R|, |B| < \epsilon n$, $|Y| < \delta n_0/2$.
\end{lm}

\noindent
{\em Proof}: 
$|R| < \epsilon n$ 
is directly obtained from 
Lemma~\ref{lm:degreebound}. 
Let $v \in V_i$ be a blue vertex 
such that a non-red edge exists in $E(W(v))$. 
From Lemma~\ref{lm:high_degree_is_red}, 
$d(W(v)) \leq \delta$. 
Thus $d(W(v))/w(W(v)) < \delta/(\delta/ \epsilon) = \epsilon$. 
This means that the average number of blue edges per a vertex is 
less than  $\epsilon$. 
Therefore $|B| <  \epsilon n$. 
%
From Lemma~\ref{lm:high_degree_is_red}, 
all edges incident to a vertex with degree higher than $\delta$ 
are red. 
From this it follows that  
the number of non-red edges in $E_k$ is at most  
$\delta |V_k|/2$. 
Thus the number of yellow edges in $E$ is also at most  $\delta |V_k|/2$. 
By considering $|V_k| < n_0$, 
we have 
$|Y| < \delta n_0/2$. \qed\\

Let $v^R_1$, $\ldots$, $v^R_{k_r}$ be the red vertices 
($k_r$ is the number of red vertices).
Let $W^B_1$, $\ldots$, $W^B_{k_b}$ be the blue components 
($k_b$ is the number of blue components). 
Let $W^Y_1$, $\ldots$, $W^Y_{k_y}$ be the yellow components 
($k_y$ is the number of yellow components). 
We consider a family of vertex subsets as 
$$
{\cal P} := \{ \{ v^R_i \} ~|~ i = 1, \ldots, k_r \} 
\cup \{ W^B_i ~|~ i = 1, \ldots, k_b \} 
\cup \{ W^Y_i ~|~ i = 1, \ldots, k_y \}. 
$$
From the definition of the function $W$ and the coloring. 
${\cal P}$ is clearly a partition of $V$.

Now we can prove Theorem~\ref{th:hyperfinite}. \\

\noindent
{\em Proof of Theorem~\ref{th:hyperfinite}}: 
If $n \leq \delta n_0/ (2 \epsilon)$, 
then the statement is clear by setting 
$t \geq \delta n_0/ (2 \epsilon)$. 
Thus, we assume that 
$n > \delta n_0/ (2 \epsilon)$. 
Let $G'$ be a graph obtained by deleting all red, blue, and yellow edges from $G$. 
From Lemma~\ref{lm:number_of_colored_edges}, 
the number of deleted edges is less than 
\begin{equation}
2 \epsilon n + \delta n_0/2 < 3 \epsilon n.
\end{equation}

Next, we will show that 
the maximum size of connected components in $G'$ is at most 
$\delta (\delta+1) /\epsilon$. 
Assume that there exists a connected component $G'(X) = (X, E_X)$  
consisting of more than $\delta (\delta+1) /\epsilon$ vertices in $G'$. 
$X$ includes no vertex $v$ with $d_G(v) > \delta$, since  
from 
Lemma~\ref{lm:high_degree_is_red} 
all edges in $E_G(v)$ are colored red. 
Moreover, there is no blue component $W \subseteq V$ such that 
$X \cap W \neq \emptyset$ and $X \backslash W \neq \emptyset$, 
as otherwise $X$ would be disconnected in $G'$ (by deleting blue edges). 

From this it follows that there is a blue or yellow component $W$ including $X$. 
Let $x \in V[T]$ be the (blue or yellow) vertex 
such that $W = W(x)$. 
If $x$ is a yellow vertex, 
then $w(v) \leq \delta / \epsilon$ 
(as otherwise $v$ would be colored blue), and $|X| \leq w(v) \leq \delta / \epsilon < \delta (\delta+1) /\epsilon$, which is a contradiction.  
Thus $x$ must be a blue vertex. 
Assume that $x \in V_h$. 
Let $Y \subseteq V_{h-1}$ be the set of children of $x$ (in $T$). 
$Y$ consists of an isolated clique or 
a double-isolated-clique in $G_{h-1}$. 
For every vertex $y \in Y$,  $d_{G_{h-1}}(y) \leq \delta$ 
(from Lemma~\ref{lm:high_degree_is_red}). 
Thus $|Y| \leq \delta+1$. 

Let $Y' \subseteq Y$ be the set of uncolored children of $x$. 
For $v \in Y'$, $w(v) \leq \delta/\epsilon$. 
Hence, 
$$
w(x) = \sum_{v \in Y'} w(v) \leq |Y'| \cdot \delta/\epsilon 
\leq |Y| \cdot \delta/\epsilon
\leq (\delta +1) \delta / \epsilon,
$$ 
which is a contradiction. 
Therefore, the maximum size of connected components in $G'$ is $\delta (\delta+1) /\epsilon$.

Thus, we have proved that $G$ is 
($\max \{ \delta n_0/(2 \epsilon), \delta (\delta+1) /\epsilon \}, 3 \epsilon$)-hyperfinite. 
Here, $\epsilon$ is an arbitrary real number in $(0,1]$, 
then by defining 
$t_{\mbox{\scriptsize \ref{th:hyperfinite}}} 
= \max \{ 3 \delta n_0 / (2 \epsilon), 3 \delta (\delta+1) /\epsilon \}$, 
$G$ is 
($t_{\mbox{\scriptsize \ref{th:hyperfinite}}}, \epsilon$)-hyperfinite 
for any $\epsilon >0$. \qed\\

%

\section{Testing Algorithm}

\subsection{Deterministic partitioning oracle}

The global partitioning algorithm of Theorem~\ref{th:hyperfinite} 
can be easily revised to run locally, 
i.e., a ``partitioning oracle'' based on this algorithm can be obtained. 
A partitioning oracle, which calculates a partition realizing hyperfiniteness locally, 
was introduced by Benjamini, et al.~\cite{BSS_MC-testable_STOC08} 
implicitly and by Hassidim, et al.~\cite{HKNO_LocalPartition_FOCS09} explicitly. 
It is a powerful tool for constructing constant-time algorithms 
for sparse graphs. It has been revised by some researchers and 
Levi and Ron's algorithm~\cite{Levi-Ron_PO_ICALP13} is the fastest to date. 
All algorithms for partitioning oracles presented to date have been randomized algorithms. 
Our algorithm, however, does not use any random valuable 
and it runs deterministically. 
That is, we call it a {\em deterministic partitioning oracle}, 
which is rigorously defined as follows\footnote{
However, since Levi and Ron's algorithm~\cite{Levi-Ron_PO_ICALP13} 
looks fast, using it may be better in practice. 
}: 

\begin{df}
${\cal O}$ is a deterministic $(t,\epsilon)$-partitioning oracle for 
a class of graphs ${\cal C}$, if, 
given query access to a graph $G = (V,E)$, 
it provides query access to a partition ${\cal P}$ of $G$. 
For a query about $v \in V$, 
${\cal O}$ returns 
${\cal P}(v)$. 
The partition has the following properties: 
(i) ${\cal P}$ is a function of $G$, $t$, and $\epsilon$. 
(It does not depend on the order of queries to ${\cal O}$.)
(ii) For every $v \in V$, 
$|{\cal P}(v)| \leq t$ and 
${\cal P}(v)$ induces a connected subgraph of $G$. 
(iii) If $G \in {\cal C}$, then 
$| \{ (u,v) \in E ~|~ {\cal P}(u) \neq {\cal P}(v) \}| \leq \epsilon |V|$. 
\end{df}

\begin{lm}\label{lm:PO}
There is a deterministic $(t_{\mbox{\scriptsize \ref{th:hyperfinite}}}, \epsilon)$-partitioning oracle ${\cal O}_{HSF}$ for HSF with $\gamma >2$ with query complexity $\delta^{O(\delta^2/\epsilon + n_0)}$ for one query. 
\end{lm}

Before giving a proof of this lemma, we introduce some notation as follows. 
A connected graph $G = (V,E)$ with a specified marked vertex $v$ is called a
{\em rooted graph}, and we sometimes say that $G$ is rooted at $v$.  
A rooted graph $G=(V,E)$ has a radius $t$, 
if every vertex in $V$ has a distance at most $t$ from the root $v$. 
Two rooted graphs are isomorphic if there is a graph isomorphism between these graphs 
that identifies the roots with each other. 
We denote by $N(d,t)$ the number of all non-isomorphic 
rooted graphs with a maximum degree of $d$ and a maximum
radius of $t$. 
For a graph $G=(V,E)$, integers $d$ and $t$, 
and a vertex $v \in V$, let $B_G(v,d,t)$ be the subgraph  
rooted at $v$ that is induced by all vertices 
of $G|d$ that are at distance $t$ or less from $v$. 
$B_G(v,d,t)$ is called a {\em $(d,t)$-disk} around $v$.  
From these definitions, the number of possible non-isomorphic 
$(d,t)$-disks is at most $N(d,t)$. \\

\noindent
{\em Proof of Lemma~\ref{lm:PO}}: 
The global algorithm of Theorem~\ref{th:hyperfinite} 
can be easily simulated locally. 
To find ${\cal P}(v)$, 
if $d(v) > \delta$, then the algorithm outputs ${\cal P}(v) := \{ v \}$. 
Otherwise, if the algorithm finds a vertex $u$ with $d(u) > \delta$ 
in the process of the local search, 
$u$ is ignored (the algorithm does not check the neighbors of $u$). 
Thus, the algorithm behaves as on the bounded-degree model. 
For any vertex $v$, 
$|{\cal P}(v)| \leq t_{\mbox{\scriptsize \ref{th:hyperfinite}}} = O(\delta^2/\epsilon)$. 
Each $u \in B_G(v,\delta,t_{\mbox{\scriptsize \ref{th:hyperfinite}}})$ 
may be included in 
${\cal P}(w)$ of $w \in B_G(u,\delta,t_{\mbox{\scriptsize \ref{th:hyperfinite}}})$. 
Then, the algorithm checks most vertices in 
$B_G(v,\delta,2t_{\mbox{\scriptsize \ref{th:hyperfinite}}}) = B_G(v,\delta,O(\delta^2/\epsilon +n_0))$, 
and thus the query complexity for one call of ${\cal P}(v)$ is at most 
$\delta^{O(\delta^2/\epsilon + n_0)}$. \qed\\

\subsection{Abstract of the algorithm}

The method of constructing a testing algorithm based on the partitioning oracle of Lemma~\ref{lm:PO} is almost the same as the method used in \cite{NS_Testable_SJCOMP13}. 
We use a distribution vector, which will be defined in Definition~\ref{df:dist_vector}, 
of rooted subgraphs consisting of at most a constant number of vertices. 

\begin{df}\label{df:dist_vector}
For a graph $G=(V,E)$ 
and integers $d$ and $t$, 
let $\mbox{\rm disk}_G(d,t)$ be the 
distribution vector of all $(d,t)$-disks of $G$, 
i.e., 
$\mbox{\rm disk}_G(d,t)$ 
is a vector of dimension $N(d,t)$. 
Each entry of $\mbox{\rm disk}_G(d,t)$ 
corresponds to some fixed rooted graph $H$, 
and counts the number of 
$(d,t)$-disks of $G|d$ 
that are isomorphic to $H$. 
Note that 
$G|d$ has $n = |V|$ different disks, 
thus the sum of entries in 
$\mbox{\rm disk}_G(d,t)$ 
is $n$. 
Let $\mbox{\rm freq}_G(d,t)$ 
be the normalized distribution, 
namely 
$\mbox{\rm freq}_G(d,t) = \mbox{\rm disk}_G(d,t)/n$. 
\end{df}

For a vector $v = (v_1, \ldots, v_r)$, 
its $l_1$-norm is 
$|| v ||_1 = \sum_{i=1}^{r} |v_i|$. 
The $l_1$-norm is also the length of the vector. 
We say that the two unit-length vectors $v$ and $u$ are 
$\epsilon$-close for $\epsilon >0$ 
if $|| v-u ||_1 \leq \epsilon$.

By using a discussion 
that is the same as in Theorem~3.1 in 
\cite{NS_Testable_SJCOMP13}, 
the following lemma is proved. 

\begin{lm}\label{lm:iff}
There exist functions 
$\lambda_{\mbox{\rm \scriptsize \ref{lm:iff}}} 
= \lambda_{\mbox{\rm \scriptsize \ref{lm:iff}}} ({\cal HFS}, \epsilon)$, 
$d_{\mbox{\rm \scriptsize \ref{lm:iff}}} 
= d_{\mbox{\rm \scriptsize \ref{lm:iff}}} ({\cal HFS}, \epsilon)$, 
$t_{\mbox{\rm \scriptsize \ref{lm:iff}}} 
= t_{\mbox{\rm \scriptsize \ref{lm:iff}}} ({\cal HFS}, \epsilon)$, 
\and 
$N_{\mbox{\rm \scriptsize \ref{lm:iff}}} 
= N_{\mbox{\rm \scriptsize \ref{lm:iff}}} ({\cal HFS}, \epsilon)$ 
such that for every $\epsilon >0$ the following holds: 
For every $G_1, G_2 \in {\cal HFS}$ on 
$n \geq N_{\mbox{\rm \scriptsize \ref{lm:iff}}}$ vertices, 
if 
$| \mbox{\rm freq}_{G_1} (d_{\mbox{\rm \scriptsize \ref{lm:iff}}}, t_{\mbox{\rm \scriptsize \ref{lm:iff}}}) 
- \mbox{\rm freq}_{G_2} (d_{\mbox{\rm \scriptsize \ref{lm:iff}}}, t_{\mbox{\rm \scriptsize \ref{lm:iff}}}) |
 \leq \lambda_{\mbox{\rm \scriptsize \ref{lm:iff}}}$, 
 then 
 $G_1$ and $G_2$ are $\epsilon$-close. \qed
\end{lm}

A sketch of the algorithm is as follows. 
Let $G=(V,E)$ be a given graph and 
$P$ be a property to test. 
First, we select some (constant) number $\ell = \ell (\epsilon)$ 
of vertices $v_i \in V$ ($i = 1, \ldots, \ell$) and 
find ${\cal P}(v_i)$ given by Theorem~\ref{th:hyperfinite}. 
This is done locally (shown by Lemma~\ref{lm:PO}).
Consider a graph 
$G' := {\cal P}(v_1) \cup \cdots \cup {\cal P}(v_{\ell})$. 
Here, $\mbox{\rm freq}_{G}(d,t)$ and $\mbox{\rm freq}_{G'}(d,t)$ 
are very close with high provability. 
Next, we calculate $\min_{G \in P}| \mbox{\rm freq}_{G'}(d,t) - \mbox{\rm freq}_{G}(d,t) |$ approximately. 
There is a problem in that the number of graphs in $P$ is generally 
infinite. However, to approximate it with a small error is 
adequate for our objective, and thus it is sufficient to compare $G'$ with 
a constant number of vectors of $\mbox{\rm freq}(d,t)$. 
(Note that calculating such a set of frequency vectors requires much time. 
However, we can say that there exists such a set.  
This means that the existence of the algorithm is assured.) 
The algorithm accepts $G$ 
if the approximate distance of 
$\min_{G \in P}| \mbox{\rm freq}_{G'}(d,t) - \mbox{\rm freq}_{G}(d,t) |$
is small enough, and otherwise it is rejected. 

The above algorithm is the same as the algorithm 
presented in \cite{NS_Testable_SJCOMP13} 
except for two points: in our model: 
(1) $G$ is not a bounded-degree graph, and 
(2) $G$ is a multigraph. 
However, these differences are trivial. 
For the first difference, it is enough to add an ignoring-large-degree-vertex process, 
i.e., if the algorithm find a vertex $v$ having a degree 
larger than $d_{\mbox{\rm \scriptsize \ref{lm:iff}}}$, 
all edges incident to $v$ are ignored.  
By adding this process, 
$G$ is regarded as $G|d_{\mbox{\rm \scriptsize \ref{lm:iff}}}$. 
This modification does not effect the result 
by Lemma~\ref{lm:degreebound}. 
For the second difference, 
the algorithm treats bounded-degree graphs as mentioned above, 
and the number of non-isomorphic multigraphs 
with $n$ vertices and degree upper bound 
$d_{\mbox{\rm \scriptsize \ref{lm:iff}}}$ is finite 
(bounded by 
$O({d_{\mbox{\rm \scriptsize \ref{lm:iff}}} }^{n^2})$). \\

\noindent
{\em Proof of Theorem~\ref{th:testable}}: 
Obtained from the above discussion. \qed

\section{Summary and future work}

We presented a natural class of multigraphs ${\cal HSF}$
representing scale-free networks, and we showed that a very wide subclass of it is hyperfinite (Theorem~\ref{th:hyperfinite}). 
By using this result, the useful result that every property is testable on the class (Theorem~\ref{th:testable}) is obtained.

${\cal HSF}$ is a class of multigraphs based on the hierarchical structure of isolated cliques.  
We may relax ``isolated cliques'' to ``isolated dense subgraphs~\cite{IsoClique_TALG09}''
and we may introduce a wider class.  
We consider such classes also to be hyperfinite.  
Finding such classes and proving their hyperfiniteness is important future work.

\section*{Acknowledgements}
We are grateful to Associate Professor Yuichi Yoshida of 
the National Institute of Informatics for his valuable suggestions. 
We also appreciate the fruitful discussions with Professor Osamu Watanabe of the Tokyo Institute of Technology and Associate Professor Yushi Uno of Osaka Prefecture University. 
We also would like to thank the Algorithms on Big Data project (ABD14) of CREST, JST, 
the ELC project (MEXT KAKENHI Grant Number 24106003), and JSPS KAKENHI Grant Numbers 24650006 and 15K11985 through which this work was partially supported.



\begin{thebibliography}{99}
%
\bibitem{AlbertBarabasi_SF_02}
R. Albert and A. -L. Barab\'asi: 
Statistical mechanics of complex networks, 
Review of Modern Physics, Vol.~74, 2002, pp.~47--97.
%
%
\bibitem{AlonFNS_Testable_Regularity_SIAMJC09}
N. Alon, E. Fischer, I. Newman, and A. Shapira:
A combinatorial characterization of the testable graph properties: it's all about regularity, 
SIAM J. Comput., Vol.~39, No.~1, 2009, pp.~143--167. 
%
%
\bibitem{BSS_MC-testable_STOC08}
I. Benjamini, O. Schramm, and A. Shapira: 
Every minor-closed property of sparse
graphs is testable, Proc. STOC 2008, ACM, 2008, pp.~393--402. 
%
\bibitem{BM_Kronecker_IPL98}
A. Bottreau, Y. M\'etivier: 
Some remarks on the Kronecker product of graphs, 
Information Processing Letters, Vol. 68, Elsevier, 1998, pp.~55--61. 
%
%
\bibitem{Broder_etal-GSinWeb-CN00}
A.Z. Broder, S.R. Kumar, F. Maghoul, R. Raghavan, S. Rajagoplalan, 
R.~Stata, A.~Tomkins and J.L.~Wiener: 
Graph structure in the Web, 
Computer Networks, Vol.~33, 2000, pp.309--320. 
%
\bibitem{CU_SF_Clique_10}
C. Cooper and R. Uehara: 
Scale free properties of random $k$-trees; 
Mathematics in Computer Science, Vol.~3, 2010, pp.~489--496. 
%
%
\bibitem{hyperfinite}
G.~Elek: 
$L^2$-spectral invariants and convergent sequence of finite graphs, 
Journal of Functional Analysis, Vol.~254, No.~10, 2008, pp.~2667--2689.
%
%
\bibitem{Gao_SF_clique_TCS09}
Y. Gao: 
The degree distribution of random $k$-trees, 
Theoretical Computer Science, Vol.~410, 2009, pp.~688--695. 
%
\bibitem{PropertyTestingLNCS10}
O. Goldreich (Ed.): 
Property Testing --- Current Research and Surveys, 
LNSC 6390, 2010. 
%
%
\bibitem{GR-STOC97}
O. Goldreich and D. Ron: 
Property testing in bounded
degree graphs: Proc. STOC 1997, 1997, pp.~406--415.
%
%
\bibitem{GGR-JACM98}
O. Goldreich, S.~Goldwasser, and D.~Ron: 
Property testing and its connection to learning and approximation: 
Journal of the ACM, Vol.~45, No.~4, July, 1998, pp.~653--750. 
%
%
\bibitem{HKNO_LocalPartition_FOCS09}
A.~Hassidim, J.~A.~Kelner, H.~N.~Nguyen, and K.~Onak:
Local graph partitions for approximation and testing, 
Proc. FOCS 2009, IEEE,  pp.~22--31.
%
\bibitem{ScalefreeTest_itohiro15_arXiv}
H.~Ito, 
Every property is testable on a natural class of scale-free multigraphs, 
arXiv: 1504.00766, Cornell University, April 6, 2015.
%
\bibitem{IsoClique_TALG09}
H.~Ito and K.~Iwama, 
Enumeration of isolated cliques and pseudo-cliques, 
ACM Transactions on Algorithms, Vol.~5, Issue~4, 
Oct. 2009, Article 40 (pp.~1--13). 
%
\bibitem{IsoClique_ESA05}
H.~Ito and K.~Iwama, and T.~Osumi: 
Linear-time enumeration of isolated cliques, 
Proc. ESA2005, LNCS, 3669, Springer, 2005, pp. 119--130.
%
\bibitem{KleinbergLawrence-web-science01}
J. Kleinberg and S. Lawrence: 
The structure of the Web, 
Science, Vol.~294, 2001, pp.~1894--1895.
%
%
\bibitem{Kusumoto-Yoshida-ICALP14}
M.~Kusumoto and Y.~Yoshida: 
Testing forest-isomorphizm in the adjacency list model, 
Proc. of ICALP2014~(1), 
LNSC 8572, 2014, pp.~763--774. 
%
%
\bibitem{Levi-Ron_PO_ICALP13}
R.~Levi and D.~Ron: 
A quasi-polynomial time partition oracle
for graphs with an excluded minor, 
Proc. ICALP 2013 (1), LNCS, 7965, Springer, 2013, 
pp.~709--720. 
(Journal version: R.~Levi and D.~Ron: 
A quasi-polynomial time partition oracle
for graphs with an excluded minor, 
ACM Transactions on Algorithms, 
Vol.~11, No.~3, 2014, Article~24 (pp.~1--13).)
%
%
\bibitem{StochKronecherG_WAW07}
M. Mahdian and Y. Xu: 
Stochastic Kronecker graphs, 
Proc. WAW 2007, LNCS, 4863, Springer, 2007, pp.~179--186. 
%
%
%
\bibitem{NS_Testable_SJCOMP13}
I. Newman and C. Sohler: 
Every property of hyperfinite graphs is testable, 
Proc. STOC 2011, ACM, 2011, pp.~675--784. 
(Journal version: 
I. Newman and C. Sohler: 
Every property of hyperfinite graphs is testable, 
SIAM~J.~Comput., Vol.~42, No.~3, 2013, pp.~1095--1112.)
%
\bibitem{Newman_SF_03}
M. E. J. Newman: The structure and function of complex networks, 
SIAM Review, Vol.~45, 2003, pp.~167--256.
%
\bibitem{Uno-Watanabe_ScaleFree}
T. Shigezumi, Y. Uno, and O. Watanabe: 
A new model for a scale-free hierarchical structure of isolated cliques, 
Journal of Graph Algorithms and Applications, Vol.~15, No.~5, 2011, pp.~661--682. 
%
\bibitem{WattsStrogatz-Nature98}
D.J. Watts and S.H. Strogatz: 
Collective dynamics of 'small-world' networks, 
Nature, Vol.~393, 1998, pp.~440--442.
%
\bibitem{ZRC_SF_clique_06} 
Z. Zhang, L. Rong, and F. Comellas: 
High-dimensional random apollonian networks, 
Physica A: Statistical Mechanics and its Applications, 
Vol.~364, 2006, 610--618.
%
%
\end{thebibliography}



\end{document}